\documentstyle[twocolumn,prb,epsfig,aps]{revtex}

\begin{document}
\twocolumn[
\hsize\textwidth\columnwidth\hsize\csname@twocolumnfalse\endcsname
\title{Magnetic  phase separation in ordered alloys}
\author{Jordi Marcos, Eduard Vives and Teresa Cast\'an}
\address{Departament d'Estructura i Constituents de la Mat\`{e}ria,
Facultat de F\'{\i}sica, \\Universitat de Barcelona, Diagonal 647,
E-08028 Barcelona, Catalonia (Spain).}
\maketitle
\begin{abstract}
We present a  lattice model to study the  equilibrium phase diagram of
ordered  alloys  with  one  magnetic  component that  exhibits  a  low
temperature  phase separation  between paramagnetic  and ferromagnetic
phases.  The model is constructed from the experimental facts observed
in   Cu$_{3-x}$AlMn$_{x}$    and   it   includes    coupling   between
configurational and magnetic degrees of freedom which are appropriated
for reproducing  the low  temperature miscibility gap.   The essential
ingredient  for the  occurrence of  such a  coexistence region  is the
development of  ferromagnetic order  induced by the  long-range atomic
order  of  the  magnetic  component.   A  comparative  study  of  both
mean-field  and Monte  Carlo  solutions is  presented.  Moreover,  the
model  may enable  the study  of  the structure  of the  ferromagnetic
domains  embedded in  the non-magnetic  matrix.  This  is  relevant in
relation to phenomena such as magnetoresistance and paramagnetism.
\end{abstract}
]
\section{Introduction}

Recently, renewed interest has been addressed to ferromagnetic ordered
alloys.  This  is because  of the unique  properties arising  from the
interplay between  elasticity, magnetism and  (configurational) atomic
order.  From the point of view of applications, the development of new
actuator materials  having very large  magnetostrains \cite{Ullako96a}
is  particularly interesting .   Also relevant  is the  possibility of
having     superparamagnetism      \cite{Yefimova87}     and     giant
magnetoresistance \cite{Yiping96} both  associated with coexistence of
magnetic domains (large magnetic particles) embedded in a non-magnetic
matrix.   This mixed  phase has  been observed,  for instance,  in the
Cu$_{3-x}$AlMn$_{x}$ Heusler alloy.

The  Heusler  alloys  are  ternary intermetallic  compounds  with  the
composition X$_2$YZ  and a low temperature $L2_1$  structure.  At high
temperatures the stable phase corresponds to a disordered bcc lattice,
also  called $A_2$  phase,  and undergoes  a two-stage  disorder-order
transition $A_2 \rightarrow B_2  \rightarrow L2_1$, as the temperature
is decreased.  Especially interesting  are the Mn-based Heusler alloys
\cite{Webster69,Webster84,Robinson97,Plogmann99},   which   exhibit  a
magnetic    moment   approximately   located    on   the    Mn   atoms
\cite{Ishikawa77}.  Among  them, the most extensively  studied are the
Ni$_2$GaMn      \cite{Planes,Castan99}     and      the     Cu$_2$AlMn
\cite{Soltys81,Bouchard75,Prado93,Obrado97,Obrado98,Obrado99a,Obrado99b}
alloys.   In both  cases, the  $L2_1$ phase  is ferromagnetic  but the
$B_2$ phase is paramagnetic.  This close relation between atomic order
and magnetic  properties has been  known to scientists for  many years
\cite{Bradley34}.   Additionally,  these  alloys exhibit  shape-memory
effects,  intimately  related to  the  structural  transition, of  the
martensitic type  \cite{Delaey91}, undergone at  low temperatures.  It
has  been suggested  that the  control of  shape-memory  properties by
application of an external magnetic field is a principle for operation
of the new class of actuator materials \cite{Ullako96b,Likhachev99}

\begin{figure}
\epsfig{file =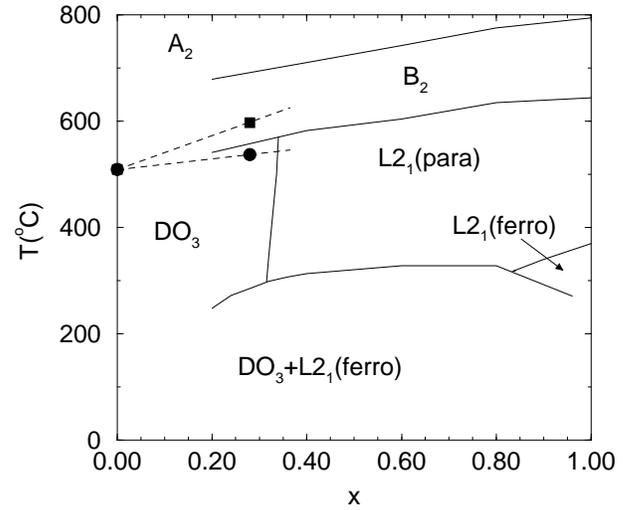, width=8cm}
\caption{Approximate experimental  phase diagram of Cu$_{3-x}$AlMn$_x$
from Ref.\ 24, 25 and 26 }
\label{FIG1}
\end{figure}

In Cu-Al-Mn the martensitic transition only exists \cite{Obrado97} for
compositions which  are very  far from the  stoichiometry (Cu$_2$AlMn)
where  the   $L2_1$  ordered  phase   is  paramagnetic  \cite{NiGaMn}.
Nevertheless, the influence of magnetism coming from Mn is revealed in
several  experiments   \cite{Experiments}.   As  well   as  the  phase
transitions mentioned above, the system exhibits, at low temperatures,
a  spinodal  decomposition  along   the  line  Cu$_3$Al  -  Cu$_2$AlMn
\cite{Soltys81,Bouchard75,Kainuma98}.  We will centre our attention on
this  two-phase region  and denote  the Cu-rich  portion of  the phase
diagram of interest in this  paper by Cu$_{3-x}$AlMn$_x$, with $0 \leq
x  \leq  1$,  .   In  Figure  \ref{FIG1}  we  show  schematically  the
corresponding phase  diagram as it  is obtained from  experiment.  The
continuous    lines   are    drawn    from   the    data   in    Ref.\
\onlinecite{Kainuma98}, whereas  the points at $x=0$  and $x=0.28$ are
from   Refs.\    \onlinecite{Obradotesi}   and   \onlinecite{Nicolaus}
respectively.

The low temperature ordered structures  for the limiting values of $x$
are  different.    The  Cu$_3$Al  binary   alloy  is  $DO_3$   at  low
temperatures,  \cite{Murray86} whereas  the Cu$_2$AlMn  is  $L2_1$ and
ferromagnetic, with a relatively high Curie temperature ($\sim 630$ K)
\cite{West56}.  The  ferromagnetism of the  $L2_1$ phase appears  as a
consequence of the atomic ordering of  the Mn atoms.  In this sense it
is  known  that properties  such  as  the  saturation magnetic  moment
depends on the degree of  order of the Mn atoms \cite{Johnston68}.  It
then  naturally  follows that  the  absence  of magnetism  (long-range
magnetic order) either  in the high temperature $B_2$-phase  or in the
low temperature  phase ($DO_3$  or $L2_1$), for  small values  of $x$,
might well be  related to the tendency for the  Mn atoms to distribute
themselves  randomly at  the different  lattice sites.   On  the other
hand,  by   increasing  the  amount   of  Mn,  for  instance   in  the
Cu$_3$AlMn$_2$   alloy,   the   resulting  magnetic   interaction   is
antiferromagnetic   \cite{Johnston68}.     Such   different   magnetic
behaviour  may  be  understood   in  terms  of  the  oscillatory  RKKY
interaction   between   the  magnetic   moments   of   the  Mn   atoms
\cite{Tajima77,Obrado99b,Vives}.

The  phase  separation or  miscibility  gap  in  Cu$_2$AlMn occurs  at
temperatures   below  $\sim   600$K   \cite{Bouchard75}  (see   Figure
\ref{FIG1})  and  gives  rise   to  a  coexistence  region  between  a
non-magnetic phase  and a  ferromagnetic phase, with  low and  high Mn
content   respectively.     The   occurrence   of   superparamagnetism
\cite{Yefimova87}   or  magnetoresistance  \cite{Yiping96}   are  both
directly related to the existence of magnetic clusters ( stable $L2_1$
domains) immersed  in the non-magnetic ($DO_3$)  matrix.  Some aspects
of this phase diagram are not totally clear.  Firstly, the persistence
of a  stable $DO_3$ phase  for small values  of $x$.  Kainuma  {\sl et
al.}\cite{Kainuma98},  by using  X-ray diffraction  measurements, have
detected an abrupt change in the intensity of the superstructure peaks
at $x \simeq  0.32$.  It should be mentioned that  this effect was not
found  in  other  earlier  studies \cite{Soltys81}.   Other  important
information,  not  yet  available,  refers  to  the  different  atomic
distributions  for  the  non-stoichiometric  $L2_1$  structure.   Some
assumptions  on this matter  will be  required in  order to  perform a
theoretical study.  Other  aspects that need to be  discussed refer to
the characteristics of the coexisting phases.  They will depend on the
location  of the  $DO_3$-$L2_1$  (according to  the  results in  Ref.\
\onlinecite{Kainuma98}) and magnetic  transition lines with respect to
the coexistence  line.  More precisely, depending  on the temperatures
at which such interphases end  on the coexistence line, the phases may
be  different in  atomic order  ($DO_3$,$L2_1$) or/and  magnetic order
(ferromagnetic, paramagnetic).  In this sense, even the coexistence of
two different  paramagnetic $L2_1$ and  $L2_1'$ phases (upper  part of
the miscibility gap  in Figure 1), with a very  similar content of Mn,
are suggested \cite{Kainuma98}.

In this  paper we present a  lattice model able to  reproduce the main
features of  the equilibrium phase  diagram in this  two-phase region.
The  details  of  the  model   will  be  derived  from  a  microscopic
description    of   the    atomic   and    magnetic    properties   of
Cu$_{3-x}$AlMn$_x$ alloys.   Nevertheless, it can be  applied to other
systems.  Practical  reasons will require several  hypotheses which in
some  cases are not  totally justified  a priori  but only  later from
agreement of obtained results  with experimental data.  This agreement
is  indicative  that the  model  captures  the  essential physics  and
provides  a starting point  for future  more exhaustive  studies.  The
model is a projection of the  ternary alloy onto a binary system, when
one of the  species is magnetic.  It is constructed  on the basis that
the main physics  of the phenomena lies on the  atomic ordering of the
magnetic  component which  moreover is  taken  to be  always the  less
abundant.    The   effective  Hamiltonian   accounts   for  a   purely
configurational  ordering energy between  first neighbouring  pairs so
that  at  low  temperatures  the  magnetic atoms  tend  to  be  second
neighbours.   Then  a simple  ferromagnetic  pair interaction  between
next-nearest neighbours  is enough to  give rise to a  low temperature
phase  separation between  a  non-magnetic phase  and a  ferromagnetic
phase that, moreover, may have different ordered structures.

It  has been  suggested \cite{Kainuma98}  that the  occurrence  of the
two-phase region in Cu$_{3-x}$AlMn$_x$  cannot be attributed to either
chemical(configurational) or magnetic ordering.   In this work, we use
a very simple microscopic model to show that the coupling between both
atomic (configurational) and magnetic  orderings is sufficient to give
rise to a decomposition between  two phases at low temperatures.  This
coupling operates in  such a way that as  the atomic ordering develops
the (indirect) exchange interactions between the atomic moments of the
magnetic particles produce a long-range ferromagnetic order.

The remainder of the paper is  organized as follows.  In section II we
introduce  the  model.   Section  III  is devoted  to  its  mean-field
solution.  In order  to better understand the nature  of the different
phases  and the  behaviour of  several measurable  quantities  we also
solve the model  by using Monte Carlo numerical  simulations.  This is
presented  in section  IV.   Finally  in section  V  we summarize  and
conclude.

\section{Model}

In the present study, our main  goal is to understand the formation of
the miscibility  gap in  Cu$_{3-x}$AlMn$_x$ along the  line $0  \leq x
\leq 1$.   The complexity  inherent to the  description of  a magnetic
ternary alloy has led us to make simplifications that we shall discuss
in  this section.   Indeed, the  quest for  reasonable simplifications
becomes  compulsory in  order  to perform  the  Monte Carlo  numerical
simulations.  Although the inclusion of too many ingredients (and thus
free  parameters)  in the  model  may  lead to  a  better  fit of  the
available data (in our case  scarce), it may hide the understanding of
the   relevant  physical  mechanism   underlying  the   phase  diagram
properties, which  we believe is  the coupling between  the long-range
configurational (chemical) ordering and the magnetism of the Mn atoms.

\begin{figure}
\epsfig{file =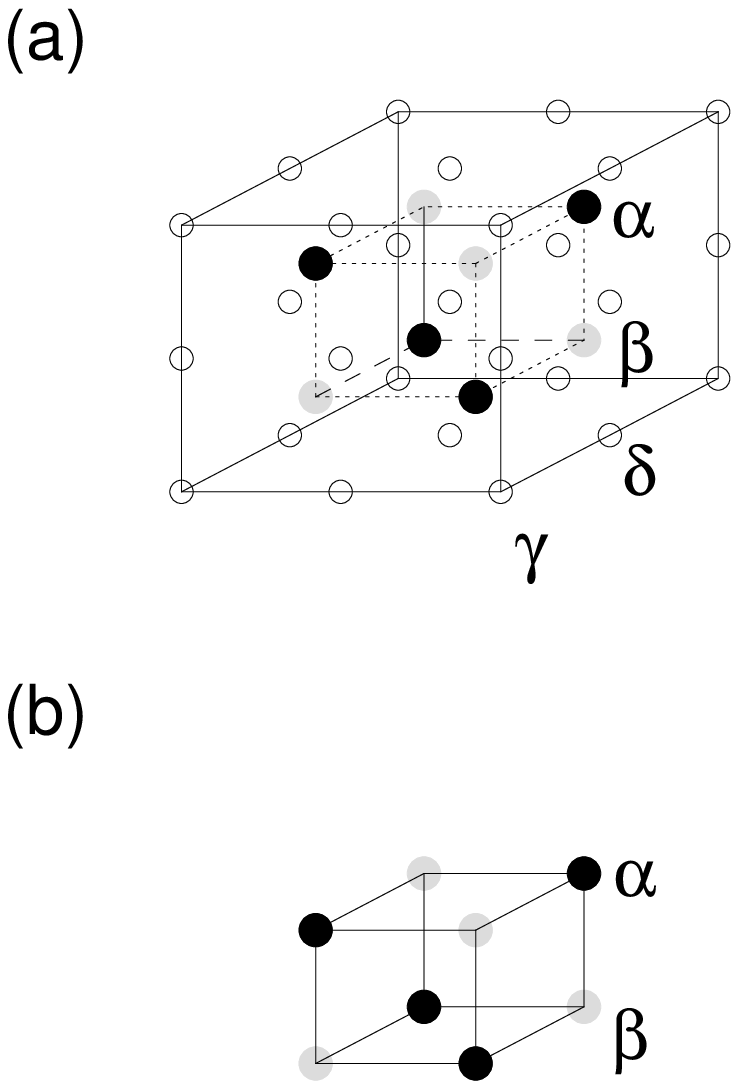, width=8cm}
\caption{(a) Structure of the $L2_1$ phase of Cu$_2$AlMn indicating the 
$\alpha$, $\beta$, $\gamma$ and $\delta$ sublattices. (b) Cell used for 
the
present model.}
\label{FIG2}
\end{figure}

The equilibrium structure of Cu$_{3-x}$AlMn$_x$ can be described as an
underlying  $bcc$  structure  formed  by  the  superposition  of  four
interpenetrated $fcc$  sublattices, named $\alpha$,  $\beta$, $\gamma$
and  $\delta$ (see  Figure  \ref{FIG2}a).  In  order  to describe  the
different  phases  of the  system  it  is  convenient to  specify  the
occupation probabilities  $p_{X}^{\Sigma}$ of the  different species X
(=Cu,Al,Mn)  in the  four different  sublattices  $\Sigma$ (=$\alpha$,
$\beta$,  $\gamma$, $\delta$).   Tables \ref{TABLE1}  and \ref{TABLE2}
summarize the occupation probabilities for the limiting $DO_3$ ($x=0$)
and $L2_1$ ($x=1$) stoichiometric  phases.  For intermediate values of
$x$ a more elaborate discussion is required.

We  start  with the  region  corresponding  to  small values  of  $x$,
$x\stackrel{\sim}{>} 0$.  Recently \cite{Kainuma98}, X-ray diffraction
experiments  shown  that  the  $DO_3$  structure  persists  above  the
coexistence region  for values of $x$  up to $0.32$.   In other words,
the  addition of  a small  amount of  Mn does  not break  the symmetry
$p_{X}^{\gamma}$=$p_{X}^{\delta}$=$p_{X}^{\alpha}$.   This  is  an  (a
priori) unexpected result, given  the different atomic environments of
these sublattices  in the $DO_3$ phase.   In any case,  it seems clear
that  entropy plays a  very important  role in  the stability  of this
homogeneous $DO_3$ phase.  From  Figure \ref{FIG1} it follows that for
$x\stackrel{\sim}{>}   0$  the   stability  is   extended   to  higher
temperatures as the value of  $x$ increases.  A natural hypothesis is,
therefore, to assume that (for low  values of $x$) the Mn atoms behave
as  impurities that  are randomly  distributed on  the  four different
sublattices.      The    corresponding     occupation    probabilities
$p_{X}^{\Sigma}$ are indicated in Table \ref{TABLE3}.

In  the $x\stackrel{\sim}{<}  1$ region  the  stable phase  is of  the
$L2_1$  type.   There  are   several  atomic  distributions  that  are
compatible with  the symmetry $p_{X}^{\gamma}$=$p_{X}^{\delta}$ $\neq$
$p_{X}^{\alpha}$ $\neq$  $p_{X}^{\beta}$.  Table \ref{TABLE4} displays
the  occupation probabilities  in  the most  straightforward case  for
which the Mn concentrates in a unique sublattice.  Alternatively, in a
more general way, one  might write the occupation probabilities (Table
\ref{TABLE5}) in terms of a  free parameter $\lambda$ ($0 \leq \lambda
\leq  1$).   Notice that  these  atomic  distributions  account for  a
continuous change  from the $DO_3$  phase ($\lambda=0$) to  the $L2_1$
phase ($\lambda>0$).

The next  step is  to introduce the  two major simplifications  of the
model:

\begin{enumerate}

\item The structures described in tables  I to V have the existence of
two n.n.  sublattices ($\gamma$  and $\delta$) in common which contain
most  of the  Cu atoms  and have  identical  occupation probabilities:
$p_{X}^{\gamma}=p_{X}^{\delta},  \;  \;  \forall X$.   Experimentally,
this symmetry  with respect to  the $\gamma$ and  $\delta$ sublattices
seems to  be satisfied for  any concentration and  temperature ranges.
From  now  on we  forget  about them  and  concentrate  on the  atomic
distribution  behaviour  on   the  other  two  remaining  sublattices,
motivated   by   the  feeling   that   the   breaking   down  of   the
$\alpha$-$\beta$ symmetry is  crucial in the ordering of  the Mn atoms
at low temperatures.   This, of course, will restrict  the validity of
our study to temperatures  below the $B_2$-$DO_3$ transition, which is
precisely the region  of interest here.  Therefore, the  model will be
defined  on  a simple  cubic  lattice  divided  into two  sublattices,
$\alpha$ and $\beta$, as illustrated in Figure \ref{FIG2}b.

\item Continuing with our assumption that the main physics lies on the
atomic  ordering  of  the  Mn  atoms,  we  shall  proceed  further  by
distinguishing between  magnetic and non-magnetic atoms  only.  In our
binary  alloy  model,  $A_{1-c}B_{c}$,  the non-magnetic  species  $A$
stands either  for Cu or Al,  whereas the magnetic  species $B$ stands
for Mn and  the composition is restricted to  $c<0.50$.  The behaviour
of the $B$  atoms on sublattices $\alpha$ and  $\beta$ can be regarded
as a simple  order-disorder transition.  For small values  of $c$ both
sublattices are equally populated  by B-atoms (behaving as impurities)
while for larger  values of $c$, B atoms occupy  preferably one of the
two sublattices.  Moreover, this behaviour depends on temperature.  As
regards  the configurational  ordering, the  model gives  rise  to two
phases only: disordered and ordered, corresponding to low ($DO_3$) and
high ($L2_1$)  content of the magnetic  species respectively.  Keeping
this  correspondence  in  mind,  in  what follows  we  shall  use  the
simplified notation D (disordered) and O (ordered).

\end{enumerate}

We  notice that  the  quantitative  study of  properties  such as  the
magnetization, susceptibility or other properties related to magnetism
(magnetoresistance, etc..)   is not our  goal here.  Rather,  we shall
focus  on  how  the  development  of  long-range  ferromagnetic  order
(resulting from  the interplay with  the atomic order)  determines the
phase diagram at low temperatures as  a function of the content of the
magnetic species  $c$.  In what  the model description  concerns, this
can  be achieved  by considering  localized Ising-like  spin variables
$s_i=\pm1$ associated with each $B$ atom.

We start from the following pair-interaction effective Hamiltonian:
\begin{eqnarray}
\label{fullham}
{\cal  H} & =  & {\cal  H}^{c} +  {\cal H}^{m}  = \\  \nonumber &  = &
\sum_{k=1}^{w}   \left   [   N_{AA}^k   \epsilon_{AA}^k   +   N_{AB}^k
\epsilon_{AB}^k+ N_{BB}^k \epsilon_{BB}^k \right ] + \\ \nonumber & +&
\sum_{k=1}^{w} \left  [ N_{B^+B^+}^{k} \epsilon_{++}^k  + N_{B^+B^-}^k
\epsilon_{+-}^k+ N_{B^-B^-}^k \epsilon_{--}^k \right ] \; ,
\end{eqnarray}
\noindent 
where ${\cal H}^{c} $ and  ${\cal H}^{m} $ are the configurational and
magnetic energy contributions respectively. The summation is performed
over  the  different   $k$-nearest-neighbour  shells  (up  to  $k=w$),
$N_{XX}^k$ is the number of $k$-th nearest-neighbour $X$-$X$ pairs and
$\epsilon_{XX}^k$  are  their  corresponding pair-interaction  energy.
Note that the  magnetic contribution only involves $B$  atoms. We have
indicated by  $B^+$ and  $B^-$ the two  possible magnetic  states.  To
preserve  the symmetry  under exchange  of  the $+$  and $-$  magnetic
states we take $\epsilon_{++}^k = \epsilon_{--}^k \; \; \forall k$.

Following standard procedures, we write Hamiltonian (\ref{fullham}) in
terms of  Ising-like variables defined  at each lattice site.   Let us
index the  sites of the  cubic lattice by $i=1,\cdots,  N$ ($N=L\times
L\times L$).   At each  lattice site $i$  we define the  following two
coupled  two-state  variables  $\sigma_i$  and  $S_i$.   The  variable
$\sigma_i=+1,-1$ represents the non-magnetic and magnetic species ($A$
and  $B$)  respectively,   then,  provided  $\sigma_i=-1$,  we  define
$S_i=+1,-1$ describing  the two possible  magnetic states of  each $B$
atom.

Considering  interactions up to  next-nearest neighbours  ($w=2$), the
configurational  energy  term   in  equation  (\ref{fullham})  can  be
written, neglecting constant terms, as:
\begin{equation}
\label{confham}
{\cal  H}^{c}  =  J_1^c   \sum_{ij}^{nn}  \sigma_i  \sigma_j  +  J_2^c
\sum_{ij}^{nnn} \sigma_i \sigma_j + E^c \sum_{i=1}^N \sigma_i \; ,
\end{equation}
where the first two summations are extended to nearest neighbour(n.n.)
and   next-nearest   neighbours   (n.n.n.)   respectively,   and   the
Hamiltonian parameters are:
\begin{equation}
J_1^c = \frac{\epsilon_{AA}^1 + \epsilon_{BB}^1 - 2 \epsilon_{AB}^1 }{4} 
\; ,
\end{equation}
\begin{equation}
J_2^c = \frac{\epsilon_{AA}^2 + \epsilon_{BB}^2 - 2 \epsilon_{AB}^2 
}{4}\; ,
\end{equation}
\noindent and
\begin{equation}
E^c=\frac{z_1}{2}   (\epsilon_{AA}^1   -\epsilon_{BB}^1)+\frac{z_2}{2}
(\epsilon_{AA}^2 -\epsilon_{BB}^2)\; ,
\end{equation}
\noindent  where $z_1=6$  and $z_2=12$  are  the number  of n.n.   and
n.n.n. of each lattice  site respectively.  In the Canonical ensemble,
the  last term  in equation  (\ref{confham}) is  just a  simple energy
shift  which  depends on  the  alloy  concentration.   As regards  the
magnetic energy term it can be rewritten as:
\begin{eqnarray}
\label{magham}
{\cal   H}^{m}   &=&   J_1^m  \sum_{ij}^{n.n.}    \frac{1-\sigma_i}{2}
\frac{1-\sigma_j}{2}    S_i    S_j    + \\ \nonumber &&   
J_2^m    \sum_{ij}^{n.n.n.}
\frac{1-\sigma_i}{2} \frac{1-\sigma_j}{2} S_i S_j+ \\ \nonumber && 
+ 4
K_1^m  \sum_{ij}^{n.n.}\frac{1-  \sigma_i}{2} \frac{1-\sigma_j}{2}+ \\
\nonumber &&  4
K_2^m \sum_{ij}^{n.n.n.} \frac{1- \sigma_i}{2} \frac{1-\sigma_j}{2},
\end{eqnarray}
\noindent where:
\begin{equation}
J_1^m  =  \frac{\epsilon_{++}^1-\epsilon_{+-}^1}{2} \;  \;  \; J_2^m  =
\frac{\epsilon_{++}^2-\epsilon_{+-}^2}{2}
\end{equation}
\begin{equation}
K_1^m = z_1\frac{\epsilon_{++}^1+\epsilon_{+-}^1}{8}  \; \; \; K_2^m =
z_2\frac{\epsilon_{++}^2+\epsilon_{+-}^2}{8}.
\end{equation}
We notice  that the latter two  terms in equation  (\ref{magham}) do not
depend on  the magnetic variables  $ \left \lbrace S_i  \right \rbrace
$.  Expanding  the   different  contributions  in  (\ref{magham})  and
ignoring  constant terms,  the  Hamiltonian  becomes:
\begin{eqnarray}
\label{full2}
 {\cal H} & =& (J_1^c + K_1^m ) \sum_{ij}^{nn}
\sigma_i \sigma_j + (J_2^c +K_2^m) \sum_{ij}^{nnn} \sigma_i \sigma_j + 
\\ \nonumber &+& 
J_1^m \sum_{ij}^{n.n.}  \frac{1-\sigma_i}{2} \frac{1-\sigma_j}{2} S_i 
S_j +\\ \nonumber &+& 
J_2^m \sum_{ij}^{n.n.n.}  \frac{1-\sigma_i}{2} \frac{1-\sigma_j}{2} S_i 
S_j.
\end{eqnarray}
The superscripts  in the  model parameters denote  its configurational
($^c$) or  magnetic ($^m$) origin, whereas the  subscripts mean first-
($_1$) or  second- ($_2$) neighbour interactions.  In  order to reduce
the number of free model parameters we set $J_1^m=0$.  Indeed, the n.n
magnetic  interaction between  $B-B$ pairs  is not  essential  for our
present purposes since we restrict  ourselves to the case in which the
ferromagnetism  develops  in   the  configurationally  ordered  phase.
Furthermore,  by  using  reduced  energy  units ${\cal  H}^*  =  {\cal
H}/(J_1^c+K_1^m)$,   we  get   the  following   {\sl   minimal}  model
Hamiltonian:
\begin{eqnarray}
\label{finalham}
{\cal H}^* &=& \sum_{ij}^{n.n.}  \sigma_i \sigma_j - K^* \sum_{ij}^{nnn}
\sigma_i  \sigma_j  - \\ \nonumber &-& J_2^* \sum_{ij}^{n.n.n.}   \frac{1-\sigma_i}{2}
\frac{1-\sigma_j}{2} S_i S_j ,
\end{eqnarray}
where the parameters are:
\begin{equation}
K^*= - \frac{J_2^c+K_2^m}{J_1^c+K_1^m},
\end{equation}
which  measures  the ordering  energy  between second-neighbour  pairs
either $A-A$, $B-B$ and $A-B$,  independently of the magnetic state of
atom $B$, and
\begin{equation}
J_2^*= - \frac{J_2^m}{J_1^c+K_1^m},
\end{equation}
which   accounts    for   the   ferromagnetic    interaction   between
second-neighbour $B-B$ pairs.

\section{Mean Field solution}
\label{meanfield}

This  section is  devoted  to  the solution  of  the model  introduced
previously  for the  A$_{1-c}$B$_{c}$ binary  alloy by  using standard
mean-field techniques  based on the  Bragg-Williams approximation.  We
denote the  occupation numbers  for each component  ($X=A,B^+,B^-$) in
each   sublattice  ($\Sigma=\alpha,\beta$)  by   $N_{X}^{\Sigma}$  and
consider the following order parameters:
\begin{equation}
\label{opx}
c = \frac{N_{B}^{\alpha +}+N_{B}^{\alpha -}+N_{B}^{\beta 
+}
+N_{B}^{\beta -}}{N} \; ,
\end{equation}
\begin{equation}
\label{opeta}
\eta  = 2  \frac{ N_{B}^{\alpha  +}+N_{B}^{\alpha  -}-N_{B}^{\beta +}-
N_{B}^{\beta -}}{N} \; ,
\end{equation}
\begin{equation}
\label{opm}
m  =   \frac{  N_{B}^{\alpha  +}+N_{B}^{\beta   +}-N_{B}^{\alpha  -}-
N_{B}^{\beta -}}{N} \; ,
\end{equation}
where $c$ ($0<c<0.5$)  is the molar fraction of  the magnetic species,
$\eta$ ($0<\eta<2c$) is the  atomic order parameter and $m$ ($0<m<2c$)
measures the magnetization of  the system.  Using standard procedures,
in the Grand Canonical formulation, we obtain the following expression
for the internal energy:
\begin{eqnarray}
\label{energy}
E & = & N J z_1 \left[ \frac{1}{2} (1-2c)^2 [1 - K^* \frac{z_2}{z_1}] -
\frac{\eta^2}{2} [1 + K^* \frac{z_2}{z_1}] \right . + \\ \nonumber && \left . \mu^* (1-2c) - J_2^* 
\frac{z_2}
{z_1} m^2 \right ] \; ,
\end{eqnarray} 
where $J=J_1^c+K_1^m$ and $\mu^*$ is the chemical potential difference
between the two species. The corresponding entropy is given by:
\begin{equation}
\label{entropy}
S = k_B \ln  \left[ \frac{N^{\alpha}!}{N_{A}^{\alpha}! N_{B^+}^{\alpha}! 
N_{B^-}^{\alpha}!} \right] \left[ \frac{N^{\beta}!}{N_{A}^{\beta}! 
N_{B^+}^{\beta}! N_{B^-}^{\beta}!} \right] \; . 
\end{equation}
Expressions (\ref{energy}) and (\ref{entropy}) produce  the following 
free energy:
\begin{eqnarray}
\label{free energy}
{\cal   F}^*  &=&   \frac{F}{NJz_1}  =   \frac{E}{NJz_1}  -   \left  (
\frac{T}{Jz_1}  \right) \left(  \frac{S}{N} \right)  = \\ \nonumber &=&
   \frac{1}{2}  (1-2c)^2   [1   -   K^*  \frac{z_2}{z_1}]   -
\frac{\eta^2}{2}  [1 +  K^* \frac{z_2}{z_1}] + \\ \nonumber &+&  \mu^*  (1-2c)  - J_2^*
\frac{z_2}  {z_1}  m^2    + \\ \nonumber  &+&  \frac{T^*}{4}  \left [
2(1-c-\frac{\eta}{2})      \ln      (1-c-\frac{\eta}{2})  + \right . \\ \nonumber &+&       2 (1-c+\frac{\eta}{2}) \ln (1 - c  + \frac{\eta}{2}) + \\ \nonumber &+& (c
+  \frac{\eta}{2}  + 2m)  \ln  (  c +  \frac{\eta}{2}  +  2m)  + \\ 
\nonumber &+& (c  +
\frac{\eta}{2} - 2m) \ln ( c + \frac{\eta}{2} - 2m) + \\ \nonumber  &+& \left . 2( c - \frac{\eta}{2}) \ln ( c - \frac{\eta}{2} ) -4c \ln 2 \right ] \; ,
\end{eqnarray}
with  $T^*=   \frac{Tk_B}{Jz_1}$  and  $J>0$.   The   free  energy  in
(\ref{free  energy}), in  the  absence of  magnetism,  reduces to  the
standard  case of  order-disorder, but  one  of the  species is  twice
degenerate.  When  magnetism is  taken into account,  model (\ref{free
energy}) exhibits  two phase transitions  respectively associated with
the  order  parameters  $\eta$  and  $m$.  We  denote  the  respective
transition temperatures  by $T_{\eta}^*$  and $T_{m}^*$. Since  we are
interested in  the case of $T_{m}^*<T_{\eta}^*$,  the model parameters
must be taken so that $(1+K^*)>J_{2}^*>0$.

\begin{figure}
\epsfig{file =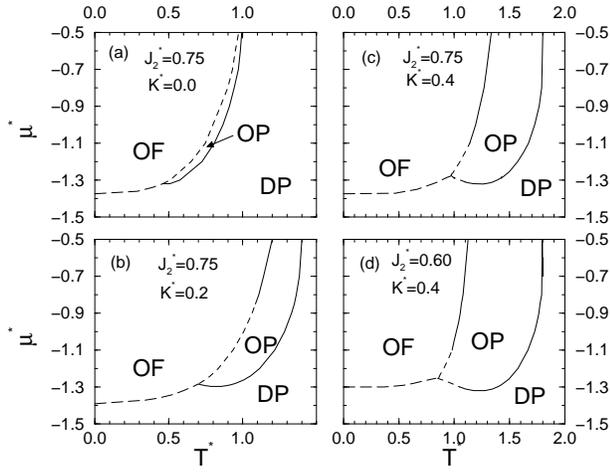, width=8cm}
\caption{Mean-field  $\mu^*$-$T^*$ diagrams  for  different values  of
model    parameters     $J_2^*$    and    $K^*$,     indicating    the
disordered-paramagnetic  phase  (DP),  the ordered-paramagnetic  phase
(OP) and the ordered-ferromagnetic  phase (OF).  Dashed lines indicate
first-order  phase  transitions   while  continuous  lines  stand  for
continuous phase transitions.}
\label{FIG3}
\end{figure}

\begin{figure}
\epsfig{file =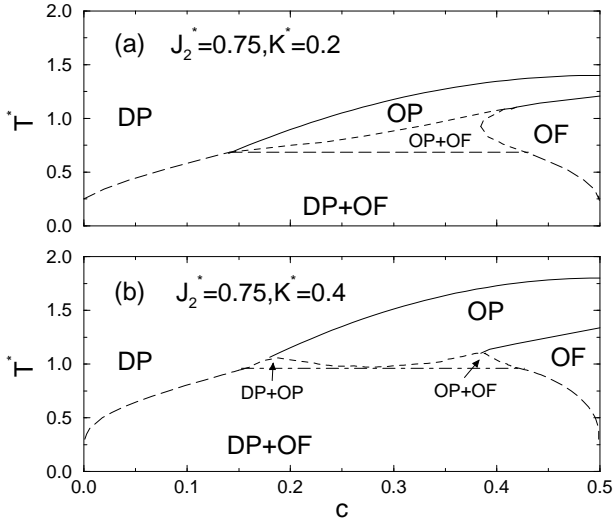, width=8cm}
\caption{Mean-Field  $c$-$T^*$  diagrams  for  two different  sets  of
values  of   model  parameters  $J_2^*$  and   $K^*$,  indicating  the
disordered-paramagnetic  phase  (DP),  the ordered-paramagnetic  phase
(OP),  the  ordered-ferromagnetic   phase  (OF)  and  the  coexistence
regions.}
\label{FIG4}
\end{figure}

The  equilibrium temperature  dependence of  the order  parameters was
obtained from direct minimization of the function (\ref{free energy}).
In figure  \ref{FIG3} we show  the $\mu^{*}-T^*$ section of  the phase
diagram for different values of $J_{2}^*=0.60,0.75$ and $K^*=0.0,0.20$
and    $0.40$.    Three    different   phases    may    appear.    The
Disordered-Paramagnetic  (DP)  phase  with  $\eta=0$  and  $m=0$,  the
Ordered-Paramagnetic (OP) phase  with $\eta \neq 0$ and  $m=0$ and the
Ordered-Ferromagnetic (OF)  phase with $\eta  \neq 0$ and $m  \neq 0$.
Both parameters $J_{2}^*$ and $K^*$  have the effect of increasing the
stability of the  ordered (OP and OF) phases.   Continuous lines stand
for second-order phase transitions,  whereas the dashed ones stand for
discontinuous phase  transitions.  The intersection  between the three
interphases corresponds  to a bicritical  point in cases (a)  and (b),
whereas for (c)  and (d) it corresponds to a  triple point.  The DP-OF
transition  is always  first order,  whereas the  other two  OP-OF and
DP-OP  may be second  or first  order.  When  the transition  is first
order, a  phase separation shows up  in the $c-T^*$  section.  This is
illustrated in  Figure \ref{FIG4} for cases (b)  and (c) corresponding
to the previous picture (Figure \ref{FIG3}).

\begin{figure}
\epsfig{file =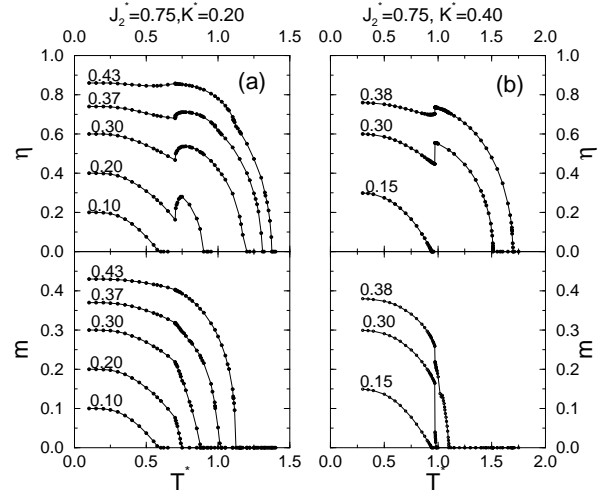, width=8cm}
\caption{Behaviour  of  the  order  parameters  $\eta$  and  $m$  with
temperature for different values of  $c$ corresponding to the same two
cases as in Fig. 4}
\label{FIG5}
\end{figure}

In  both cases  of  Figure  \ref{FIG4} a  phase  separation between  a
non-magnetic (paramagnetic) and a ferromagnetic phases exists.  At low
temperatures the coexisting phases (DP+OF) are also different in their
atomic  ordered structure,  whereas at  moderate  temperatures (OP+OF)
both  exhibit the  same atomic  structure.  Besides,  for case  (b) (a
larger  value  of  $K^*$)  a  phase  separation  (DP+OP)  between  two
non-magnetic phases  appears.  In  this case, there  exists a  line of
triple points  (horizontal dot-dashed line).  In  Figure \ref{FIG5} we
show the  corresponding temperature behaviour of  the order parameters
$\eta$  and $m$  for different  values of  the composition  $c$.  This
information  is obtained  from  the calculations  presented in  Figure
\ref{FIG4} taking into  account the fact that in  the phase separation
region the system is  heterogeneous and that at constant concentration
both  the characteristics  and  the amount  of  the coexisting  phases
change with temperature.  It is  noticeable that in both cases the two
order  parameters ($\eta$  and  $m$)  exhibit an  anomaly  at a  given
temperature,  (a) $T^*  \sim 0.70$  and  (b) $T^*  \sim 0.97$.   These
temperatures  correspond  to  the  bicritical  and  the  triple  point
discussed in  Figure \ref{FIG3}.  When crossing the  triple point line
(case (b)), the anomaly is accompanied by a discontinuity in the order
parameters.


\section{Monte Carlo simulation}
\label{montecarlo}

Monte Carlo simulations of  model (\ref{finalham}) have been performed
in order to study the  role of fluctuations.  Starting from an initial
(arbitrary)  configuration, the subsequent  microscopic configurations
are  generated by using  the standard  Metropolis algorithm.   We have
focussed on  two cases.  First, on the  stoichiometric alloy ($c=0.5$)
for different  values of the parameters $K^*$  and $J_2^*$.  Secondly,
we have  fixed $K^*=0.4$  and $J_2^*=0.6$ and  have studied  the phase
diagram as a function of $c$ and $T^*$.

\subsection{Simulation details}

The  main results  were obtained  on a  simple cubic  lattice  of size
$L=16$ ($N= L \times L \times L =4096$). Moreover, a certain number of
simulations with $L=24$  and $L=32$ were also carried  out in order to
study  finite-size  effects  and  to obtain  illustrative  real  space
snapshots of  the system. Energy and order  parameter fluctuations are
measured according to the following definitions:
\begin{equation}
{\cal C} = \frac{1}{N T^{*2}} \left ( \langle H^{*2} \rangle - \langle
H^* \rangle ^2 \right )
\label{specific}
\end{equation}
\begin{equation}
\chi_{\eta} =  \frac{1}{T^*} \left ( \langle \eta^2  \rangle - \langle
\eta \rangle ^2 \right )
\end{equation}
\begin{equation}
\chi_{m}  = \frac{1}{T^*}  \left (  \langle  m^2 \rangle  - \langle  m
\rangle ^2 \right ).
\end{equation}
The brackets  stand for  Monte Carlo (MC)  averages, performed  over a
large number of uncorrelated configurations after the equilibration of
the system. In  order to find the phase  diagram, the transition lines
were located from the positions  of the peaks of the above quantities.
In   many   cases   equilibration   was   checked   by   testing   the
fluctuation-dissipation  theorem,  i.e.:
\begin{equation}
\label{derivatives}
{\cal C} = \frac {1}{N} \frac{d \langle H^{*} \rangle }{d T^*} .
\end{equation}
Two kinds of numerical simulation experiments have been performed:

\begin{enumerate}

\item  Grand  Canonical simulations.   The  simulations  in the  Grand
Canonical   Ensemble   have   the   advantage   of   allowing   faster
equilibration.  The alloy concentration is not fixed and an additional
term  taking  into  account  the  effect  of  the  chemical  potential
difference between both species is needed in this case. Formally, this
is   done   by   a   Legendre  transformation   of   the   Hamiltonian
(\ref{finalham}). This yields:

\begin{eqnarray}
{\cal H}^* &=& \sum_{ij}^{n.n.}  \sigma_i \sigma_j - K^* \sum_{ij}^{nnn}
\sigma_i  \sigma_j  - \\ \nonumber &-&  J_2^* \sum_{ij}^{n.n.n.}   \frac{1-\sigma_i}{2}
\frac{1-\sigma_j}{2} S_i S_j + \mu^* \sum_{i=1}^N \sigma_i .
\end{eqnarray}

Starting  from an  (arbitrary)  initial configuration,  the system  at
constant $T^*$  and $\mu^*$, evolves  towards equilibrium by  means of
Glauber excitations  proposed in both variables,  $\sigma_i$ and $S_i$
independently.  The unit  of time MCS (a Monte  Carlo step) is defined
as  $N$ independent  proposals  of each  kind  of flip  on a  randomly
selected  lattice site.   Typically  the averages  are performed  over
$1500$ configurations, taken every $20$ MCS and discarding the initial
$5000$  MCS  for  equilibration.   The  regions  of  phase  separation
correspond to unreachable regions in the $c-T^*$ phase diagram.

\item  Canonical  simulations.    In  these  simulations  the  Glauber
excitations  are proposed  in the  magnetic variable  $S_i$  only.  In
order to preserve the  alloy composition $c$, the variables $\sigma_i$
evolve according to the  Kawasaki exchange dynamics. The equilibration
process is much slower in this  case and the system may get trapped in
metastable configurations.   To get rid of such  configurations, it is
convenient  to allow  a certain  fraction ($q$)  of  exchanges between
n.n.n.  atoms. Then, a MCS is in this case defined as $N$ proposals of
$S_i$ flips,  $N(1-q)$ proposals of n.n. exchanges  and $Nq$ proposals
of n.n.n. exchanges.   We have studied the effect  of different values
of $q$ and found that $q \sim 0.2$ is enough to reach equilibrium in a
reasonable  time.   Typically   averages  are  performed  over  $3500$
configurations,  taken  every $50$  MCS,  after  discarding the  first
$25000$ MCS for  equilibration. In the region of  phase separation the
simulated  system evolves to  an inhomogeneous  ``slab'' configuration
with a flat interface.  Because  of finite-size effects, the energy of
such configurations  is very much dominated by  the interfacial energy
and should be carefully analyzed. In spite of the long times needed to
get reliable  results, the simulations  in the Canonical  ensemble are
very  useful  here  since  they  provide  information  concerning  the
structure of the domains in the coexistence region.
\end{enumerate}

\subsection{Monte Carlo Results}

\begin{figure}
\epsfig{file =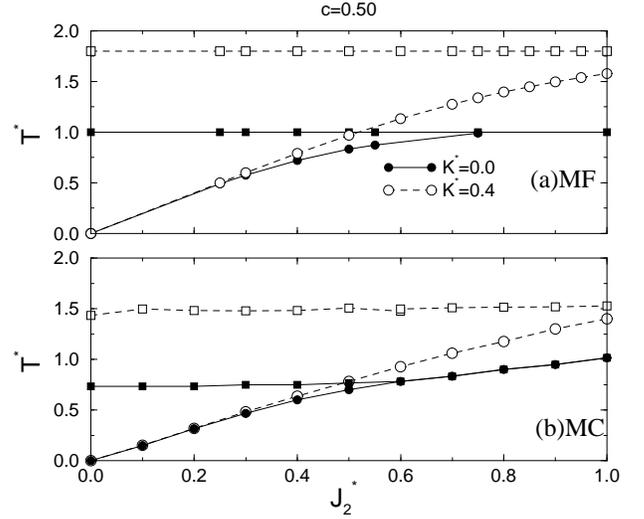, width=8cm}
\caption{Dependence   of   the   transition  temperatures   $T_\eta^*$
($\Box$, \protect\rule[0.1mm]{2mm}{2mm})   and  $T_m^*$   ($\circ$,$\bullet$)  with
$J_2^*$  for the  stoichiometric compound  ($c=0.5$),  from Mean-Field
calculations  (a)  and  Monte  Carlo  simulations  (b).  Open  symbols
correspond to $K^*=0.4$ and filled symbols to $K^*=0.0$.}
\label{FIG6}
\end{figure}

We start  by presenting the  transition temperatures as a  function of
the model parameters for the case of the stoichiometric alloy $c=0.5$.
This  is  shown  in  the  lower  part of  Figure  \ref{FIG6}  (b).   A
comparative look  of both  mean-field (a) and  MC results  (b) reveals
that  both  solutions  render  the  same  qualitative  behaviour.  The
fluctuations (taken into account in the Monte Carlo solution) have the
effect  of increasing  the stability  of the  disordered, paramagnetic
phases so that  the overall transition temperatures are  lower than in
the mean-field solution.

\begin{figure}
\epsfig{file =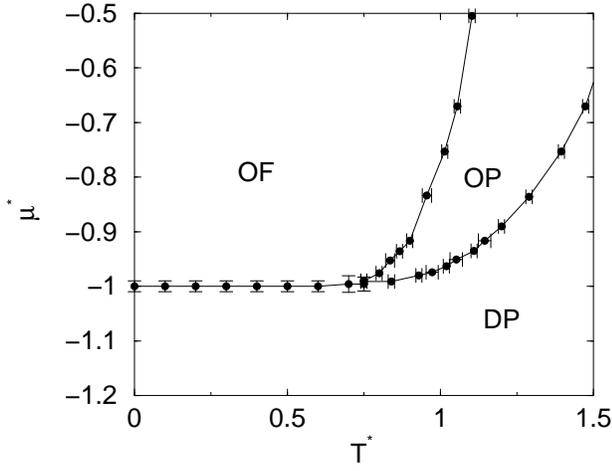, width=8cm}
\caption{Phase diagram  $\mu^*$-$T^*$ for $J_2^*=0.60$  and $K^*=0.40$
obtained  from   Monte  Carlo  simulations  in   the  Grand  Canonical
ensemble. Lines are guides to the eye.}
\label{FIG7}
\end{figure}

Figure \ref{FIG7} shows the  $\mu^*-T^*$ section of the phase diagram,
drawn from  the Grand Canonical simulations,  with $L=16$, $J_2^*=0.6$
and $K^*=0.4$.  We notice that the model parameters are those of Fig.\
\ref{FIG3}(d).   It  follows   that  both  numerical  simulations  and
mean-field techniques render the  same qualitative phase diagram.  The
only remark that comes out is  the smearing out of the re-entrant (OP)
phase in the  MC solution, due to the  fluctuations.  The available MC
data  does not  allow for  a  conclusive determination  of the  nature
(first or second-order) of the transitions.

In order  to compare  data with experiment,  the study of  the $c-T^*$
section of the phase diagram is  essential. It turns out to be a tough
task  because   of  the   finite-size  effects.   In   particular,  to
definitively  resolve the coexistence  region, one  needs to  use very
large linear system sizes.

\begin{figure}
\epsfig{file =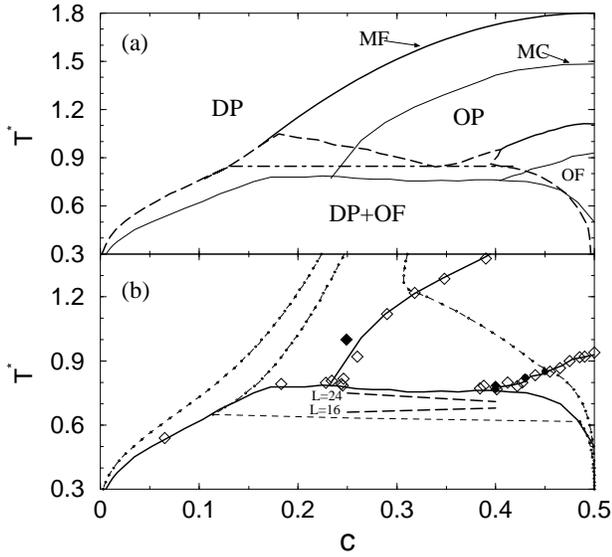, width=8cm}
\caption{Phase  diagram   $c-T^*$  for  $J_2^*=0.60$   and  $K^*=0.40$
obtained from mean-field calculations  (a) and Monte Carlo simulations
(b).   Thick   lines  indicate   the  phase  boundaries   between  the
homogeneous phases  and the coexistence regions.  The  thin lines with
dots  correspond  to  Gran  Canonical  MC runs  at  constant  chemical
potential, diamonds indicate the  positions of the specific heat peaks
from  Gran Canonical  (open diamonds)  and canonical  (black diamonds)
simulations.   Thick  dashed  lines  in  (b) are  estimations  of  the
coexistence region boundary from Canonical simulations with $L=16$ and
$L=24$.  The  continuous thin  lines in (a)  indicate the  Monte Carlo
phase boundaries for comparison.}
\label{FIG8}
\end{figure}

Figure  \ref{FIG8} shows  the $c-T^*$  phase diagram  corresponding to
$J_2^*=0.6$  and $K^*=0.4$.   In Fig.\  \ref{FIG8}a  we simultaneously
show the mean-field and the  MC solutions.  One observes that the main
trends of both phase diagrams are the same.  For practical reasons, we
show the MC  solution in more detail in Fig.\  \ref{FIG8}b . The phase
transition lines and  the limits of the coexistence  region, have been
located from the  peaks observed in the specific  heat $\cal C$.  This
criterion  has  been  followed  in  both  the  Grand  Canonical  (open
diamonds) and  Canonical (black diamonds) simulations .   In the Grand
Canonical   simulations  the   coexistence  region   is   revealed  by
unreachable zones in the  $\langle c \rangle -T^*$ diagram accompanied
by flat  steps in the curves  of constant $\mu^*$  (three examples are
depicted by small dots joined by a thin line).
 
When  comparing the  results  corresponding to  the  same system  size
obtained from  both the Canonical and the  Grand Canonical simulations
we  see  that in  the  former the  coexistence  line  occurs at  lower
temperatures.  This is  due to finite-size effects that  have a strong
influence on  the stabilization of the mixed  phase configurations. In
this sense we have checked that when the system size is increased this
effect  is  corrected  and  the  phase  separation  occurs  at  higher
temperatures.   To   illustrate  this,  we  have   plotted  in  Figure
\ref{FIG8} (with thick dashed lines) the upper part of the coexistence
line obtained from Canonical  simulations, for two different values of
the system size ($L$= 16 and $L$=24), as indicated.

\begin{figure}
\epsfig{file =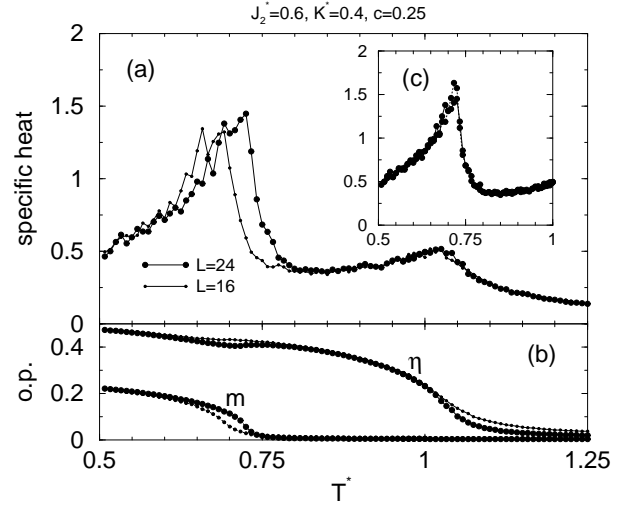, width=8cm}
\caption{Specific heat $\cal C$ and order parameters $m$ and $\eta$ as
a  function  of  temperature  $T^*$  for  $J_2^*=0.6$,  $K^*=0.4$  and
$c=0.25$ obtained  from canonical MC simulations. Data  for $L=16$ and
$L=24$ are shown with thin and thick lines respectively. The inset (c)
shows the  specific heat computed from  fluctuations (continuous line)
as well as from the derivative of the average energy (dashed line).}
\label{FIG9}
\end{figure}

The same  effect appears when  studying the specific heat.   In Figure
\ref{FIG9}  we show  the temperature  behaviour of  the  specific heat
$\cal C$ (a) together with the order parameters $m$ and $\eta$ (b) for
$J_2^*=0.6$, $K^*=0.4$ and $c=0.25$  as obtained from the canonical MC
simulations.  Data  shown correspond to  $L=16$ and $L=24$.  Note that
the peak  corresponding to  the phase separation  shows a  much larger
dependence on  $L$ than the  peak corresponding to  the order-disorder
transition.  The inset  (c) shows the specific heat  computed from the
energy fluctuations (equation  \ref{specific}) and from the derivative
of  the  average energy  (equation  \ref{derivatives}). The  agreement
ensures that the equilibration times considered are long enough.

In spite  of the difficulties described above,  which certainly hinder
the location of the boundaries,  the phase diagram presented in Figure
\ref{FIG8} is essentially similar to that obtained experimentally (see
Figure \ref{FIG1}) at least at moderate and low temperatures. The lack
of resolution in  the results makes it impossible  to conclude whether
or not the MC results render a  line of triple points as occurs in the
mean-field  solution (lower  part of  \ref{FIG4}).  Unfortunately, the
existing  experimental data  do not  provide new  information  on this
point.  We  suggest that more  experiments are needed.   Provided that
the  experimental phase  diagram is  sufficiently well  resolved, fine
tuning of the parameters $J_2^*$  and $K^*$ (even $J_1^*$) would allow
the matching of more details.

\begin{figure}
\epsfig{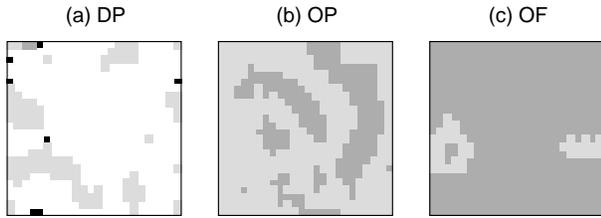}
\caption{Snapshots  of the system  configuration corresponding  to the
three  different  phases: (a)  Disordered-Paramagnetic  (DP) phase  at
$T^*=1.2$  and   $c=0.2$,  (b)  Ordered-Paramagnetic   (OP)  phase  at
$T^*=1.0$  and $c=0.32$  and (c)  Ordered-Ferromagnetic (OF)  phase at
$T^*=0.79$  and $c=0.45$.  The shading identify locally  the different
phases according to  the short-range order parameters  as explained in
the text.}
\label{FIG10}
\end{figure}

Besides the  determination of the phase diagram  and the fluctuations,
the  MC  simulations have  the  possibility  of  providing real  space
snapshots of the system configuration.  Figure \ref{FIG10} shows a two
dimensional section of the  simulated system with $L=24$ for different
homogeneous equilibrium  phases corresponding to the  phase diagram in
Figure \ref{FIG8}.  Case (a) corresponds  to the DP phase with $c=0.2$
and $T^*=1.2$, (b) to the OP phase with $c=0.32$ and $T^*=1.0$ and (c)
to the OF  phase with $c=0.45$ and $T^*=0.79$.   The assignment of the
different  colours has been  done by  measuring the  short-range order
parameters  in a cell  of size  $5\times5 \times  5$ centered  at each
lattice  site  of a  certain  two-dimensional  horizontal  cut of  the
original  system.  When  the  values of  the  local magnetization  $m$
and/or local order parameter  $\eta$ are above $0.2$ the corresponding
lattice site is  considered to belong to a  ferromagnetic and/or to an
atomically  ordered phase  respectively.  White,  light gray  and dark
gray  indicate DP,  OP and  OF  regions.  Black  corresponds to  local
disordered ferromagnetic regions which do not correspond to any stable
phase.   These  appear  because  the  fluctuations  become  both  more
probable  and important  with temperature  in the  homogeneous phases.
Actually,  the  three snapshots  correspond  to  a  time evolution  of
$2\cdot 10^5$  MCS, when  the average values  of the  long-range order
parameters  are perfectly equilibrated.   Thus, the  curved interfaces
reveal that the fluctuations evolve with time and appear and disappear
very quickly.

\begin{figure}
\epsfig{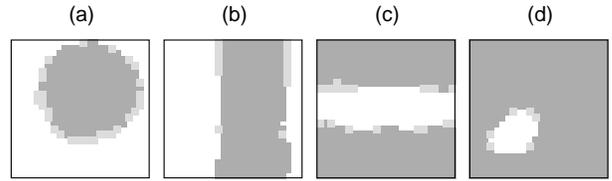}
\caption{Snapshots  of  the system  configuration  in the  coexistence
region for $T^*=0.5$ and $c= 0.1$  (a), $c= 0.2$ (b), $c= 0.3$ (c) and
$c=0.45$ (d).}
\label{FIG11}
\end{figure}

In Figure  \ref{FIG11} we show  snapshots of the  system configuration
inside  the  coexistence  region.   The four  pictures  correspond  to
$T^*=0.5$ and to different values of the composition; (a) $c=0.1$, (b)
$c=0.2$,   (c)  $c=0.3$  and   (d)  $c=0.45$.    Note  that   for  low
concentration of the magnetic component  (a), the OF phase consists of
ferromagnetic bubbles  inside the DP  matrix as expected.   For larger
values of $c$  the ferromagnetic bubbles transform into  rods or slabs
(b).  This is an artifact of finite-size effects that makes the system
decrease the  interfacial energy by  taking advantage of  the periodic
boundary conditions.  Cases  (c) and (d) are symmetric  to (b) and (a)
respectively.   Given   the  large  value   of  $c$,  the   matrix  is
ferromagnetic and the domains paramagnetic.

It  is known  that the  shape  and the  size of  the magnetic  bubbles
embedded into the non-magnetic matrix is crucial for the occurrence of
magnetoresistance. In the light of the present results, we believe the
present model is suitable  for determining the optimum characteristics
of such domains.  Along these lines, the study of  the kinetics of the
domain  growth  after quenches  from  high  temperature should  supply
useful information. This will be the subject of future work.

\section{Conclusion}

By using a simple lattice model we have shown that the magnetism of an
ordered  alloy may  give rise  to a  low temperature  phase separation
between a ferromagnetic phase  and a paramagnetic phase. The existence
of  this mixed  phase is  relevant in  relation to  the  occurrence of
phenomena such as paramagnetism and magnetoresistance.

This  study   has  been  motivated   by  the  behaviour   observed  in
Cu$_{3-x}$AlMn$_x$.   Nevertheless,  the   strategy  followed  in  the
construction of the model should apply to other alloys. In particular,
to those in which the ferromagnetism is induced by the configurational
ordering of  the magnetic atoms, as occurs  in Cu$_{3-x}$AlMn$_x$. Our
main conclusion is that this  interplay between both kind of orderings
is enough to produce the magnetic phase separation.  We should mention
that other effects such as elasticity due to the different atomic size
of the elements may affect the  final phase diagram.  In spite of this
and in view of the present results it is clear that the model captures
the essential  ingredients and makes it an  appropriate starting point
for future dynamical studies of the kinetics of formation of the mixed
phase after a suitable thermal quench.

\section*{Acknowledgements}

We acknowledge  fruitful discussions with Antoni  Planes.  The authors
also   acknowledge  financial  support   from  CICyT   project  number
MAT98-0315.   J.M.   acknowledges  financial support  from  Direcci\'o
General de Recerca (Catalonia).

\newpage

\begin{table}
\begin{displaymath}
\begin{array}{c|cccc}
 & \alpha & \beta & \gamma & \delta \\
\hline
Cu & 0 & 1 & 1 & 1 \\
Al & 1 & 0 & 0 & 0 \\
Mn & 0 & 0 & 0 & 0 \\
\end{array}
\end{displaymath}
\caption{Occupation   probabilities  for   the   stoichiometric  $DO_3$
structure of Cu$_3$Al.}
\label{TABLE1}
\end{table}

\begin{table}
\begin{displaymath}
\begin{array}{c|cccc}
 & \alpha  &  \beta  &  \gamma  &  \delta  \\
\hline
 Cu  &  0  &  0  &  1  &  1  \\
 Al  &  1  &  0  &  0  &  0  \\
 Mn  &  0  &  1  &  0  &  0  \\ 
\end{array}
\end{displaymath}
\caption{Occupation   probabilities  for   the   stoichiometric   $L2_1$ 
structure of  Cu$_2$AlMn .}
\label{TABLE2}
\end{table}

\begin{table}
\begin{displaymath}
\begin{array}{c|cccc}
 &   \alpha   &   \beta   &   \gamma   &  \delta   \\  
\hline 
 Cu  &  0&  1-\frac{x}{3}  &  1-\frac{x}{3}   &  1-\frac{x}{3}   \\  
 Al   &  1-\frac{x}{4}  &  \frac{x}{12}  &  \frac{x}{12}  &  
\frac{x}{12}    \\
 Mn  &  \frac{x}{4}  &   \frac{x}{4}  &   \frac{x}{4}  &  \frac{x}{4}    
\\ 
\end{array}
\end{displaymath}
\caption{Guessed  occupation probabilities for  the non-stoichiometric
$DO_3$ structure with $x \stackrel{\sim}{>}0$ composition.}
\label{TABLE3}
\end{table}

\begin{table}
\begin{displaymath}
\begin{array}{c|cccc}
 &   \alpha   &   \beta   &   \gamma   &  \delta   \\  
\hline  
 Cu   & 0 &  1-x  &  1   &  1   \\  
 Al  & 1  &  0  &  0  &  0   \\
 Mn  &   0  &  x  &  0  &  0  \\ 
\end{array}
\end{displaymath}
\caption{Simplest occupation  probabilities for the non-stoichiometric
$L2_1$ structure with $x \stackrel{\sim}{<}1$ composition.}
\label{TABLE4}
\end{table}

\begin{table}
\begin{displaymath}
\begin{array}{c|cccc}
 &  \alpha   &  \beta  &  \gamma  &   \delta  \\  \hline  Cu   &  0  &
 1-\frac{x}{3}(1+2\lambda) & 1-\frac{x}{3} (1-\lambda) & 1-\frac{x}{3}
 (1-\lambda)  \\   Al  &  1-\frac{x}{4}   (1-\lambda)  &  \frac{x}{12}
 (1-\lambda) & \frac{x}{12}  (1-\lambda) & \frac{x}{12} (1-\lambda) \\
 Mn  &   \frac{x}{4}  (1-\lambda)   &  \frac{x}{4}  (1+3   \lambda)  &
 \frac{x}{4} (1-\lambda) & \frac{x}{4} (1-\lambda) \\
\end{array}
\end{displaymath}
\caption{Guessed  occupation probabilities for  the non-stoichiometric
Cu$_{3-x}$ AlMn$_x$  alloys  that include the   $DO_3$  structure ( 
$\lambda=0$ )
and the $L2_1$ structure ($\lambda \neq 0$)}
\label{TABLE5}
\end{table}

\newpage

\newpage

\newpage

\begin{references}
\bibitem{NiGaMn}  In the case  of $Ni_2MnGa$,  the cubic  $L2_1$ phase
which transforms martensitically is ferromagnetic.
\end{references}
\end{document}